\documentclass{llncs}

\usepackage[utf8]{inputenc}
\usepackage{graphicx}
\usepackage{fancyhdr}

\usepackage[labelformat=simple]{subcaption}

\usepackage{float}
\usepackage{cmap}
\usepackage{cleveref}
\usepackage{xcolor}

\usepackage{listings}

\usepackage{textcomp}
\definecolor{keyword}{HTML}{2771a3}
\definecolor{pattern}{HTML}{b53c2f}
\definecolor{string}{HTML}{be681c}
\definecolor{relation}{HTML}{7e4894}
\definecolor{variable}{HTML}{107762}
\definecolor{comment}{HTML}{8d9094}

\lstdefinelanguage{cypher}
{
	morekeywords={
		MATCH, OPTIONAL, WHERE, NOT, AND, OR, XOR, RETURN, DISTINCT, ORDER, BY, ASC, ASCENDING, DESC, DESCENDING, UNWIND, AS, UNION, WITH, ALL, CREATE, DELETE, DETACH, REMOVE, SET, MERGE, SET, SKIP, LIMIT, IN,
		INDEX, DROP, UNIQUE, CONSTRAINT, EXPLAIN, PROFILE, START, CASE,
		GROUP, HAVING,
	},
	sensitive=true,
	morecomment=[l]{//},
	morecomment=[s]{/*}{*/},
	morestring=[b]{"},
}

\newcommand{\listingcypher}[2]{
	\lstset{
		language=Cypher
	}
	\lstinputlisting[label=lst:#1, caption=#2.]{adbis-examples/#1.cypher}
	\vspace{-5ex}
}

\newcommand{\mycdots}{\cdot\!\cdot\!\cdot}
\usepackage{csquotes}
\usepackage{multirow}
\usepackage{array}
\usepackage{enumitem}
\usepackage{todonotes}
\presetkeys{todonotes}{inline}{}
\usepackage[linesnumbered,lined,boxed,commentsnumbered]{algorithm2e}
\SetKwProg{Fn}{Function}{}{}
\usepackage{adjustbox}
\usepackage{stmaryrd}
\usepackage{autobreak}

\usepackage[
	bookmarks=false,
	breaklinks=true,
	pdfpagelayout=SinglePage,
	pdfstartview=Fit,
	bookmarks,
	bookmarksopen,
	bookmarksdepth=3,
	hidelinks,
]{hyperref}
\usepackage[all]{hypcap}

\newcommand{\viatraquery}{\mbox{\textsc{Viatra} Query}\xspace}
\newcommand{\vql}{\mbox{\textsc{Viatra} Query Language}\xspace}

\newcommand{\saphana}{SAP \mbox{HANA}\xspace}

\newcommand{\opencypher}{\mbox{openCypher}\xspace}

\newcommand{\sparql}{\mbox{SPARQL}\xspace}
\newcommand{\rga}{relational graph algebra\xspace}
\newcommand{\RGA}{Relational Graph Algebra\xspace}

\newcommand{\eg}{e.g.\xspace}
\newcommand{\ie}{i.e.\xspace}

\newcommand{\operatortext}[1]{#1\xspace}

\newcommand{\getverticestext}{\operatortext{get-vertices}}

\newcommand{\expandbothtext}{\operatortext{expand-both}}
\newcommand{\expandintext}{\operatortext{expand-in}}
\newcommand{\expandouttext}{\operatortext{expand-out}}
\newcommand{\projectiontext}{\operatortext{projection}}
\newcommand{\selectiontext}{\operatortext{selection}}
\newcommand{\alldifferenttext}{\operatortext{all-different}}
\newcommand{\duplicateeliminationtext}{\operatortext{duplicate-elimination}}
\newcommand{\sorttext}{\operatortext{sorting}}
\newcommand{\groupingtext}{\operatortext{grouping}}
\newcommand{\toptext}{\operatortext{top}}
\newcommand{\unwindtext}{\operatortext{unwind}}
\newcommand{\baguniontext}{\operatortext{bag union}}

\newcommand{\cartesianproducttext}{\operatortext{Cartesian product}}
\newcommand{\leftouterjointext}{\operatortext{left outer join}}
\newcommand{\jointext}{\operatortext{natural join}}
\newcommand{\uniontext}{\operatortext{union}}

\newcommand{\append}{\,\|\,}

\newcommand{\remove}{\setminus}

\newcommand{\breakable}[2][c]{%
	\begin{tabular}[#1]{@{}l@{}}#2\end{tabular}}

\newcommand{\vertexlabels}{L}
\newcommand{\edgelabels}{T}

\newcommand{\verticestoedges}{\mathit{st}}

\newcommand{\vertexproperties}{P_v}
\newcommand{\edgeproperties}{P_e}

\newcommand{\vertexlabelfunction}{\mathcal{L}}
\newcommand{\edgelabelfunction}{\mathcal{T}}

\newcommand{\dom}[1]{\mathrm{dom}(#1)}
\newcommand{\schema}[1]{\mathrm{sch}(\mathit{#1})}

\newcommand{\mylst}[1]{\lstinline{#1}}
\newcommand{\mylstc}[1]{\mbox{\lstinline{#1},}}

\usepackage{ifxetex}
\usepackage{ifluatex}
\newif\ifxetexorluatex % a new conditional starts as false
\ifnum 0\ifxetex 1\fi\ifluatex 1\fi>0
   \xetexorluatextrue
\fi

\ifxetexorluatex
  \usepackage{fontspec}
\else
  \usepackage[T1]{fontenc}
  \usepackage[utf8]{inputenc}
\fi

\usepackage{amsmath}
\usepackage{amssymb}
\usepackage{etoolbox}
\usepackage{xspace}
\newtoggle{textualoperators}

\usepackage{tikz}
\usepackage{forest}

\tikzset{every node/.style={draw}}

\newcommand{\relnull}{\mathsf{NULL}}

\newcommand{\assign}{\rightarrow}

\newcommand{\asc}{\uparrow}
\newcommand{\desc}{\downarrow}

\newcommand{\tuple}[1]{\langle #1 \rangle}

\newcommand{\literal}[1]{\mathsf{#1}}
\newcommand{\atom}[1]{\mathsf{#1}}

\newcommand{\colonseparator}{:}

\newcommand{\var}[1]{\mathtt{#1}}
\newcommand{\edgevariable}[2]{\var{#1}\ifstrempty{#2}{}{\colonseparator{\atom{#2}}}}
\newcommand{\vertexvariable}[2]{(\var{#1}\ifstrempty{#2}{}{\colonseparator{\atom{#2}}})}

\def\ojoin{\setbox0=\hbox{$\bowtie$}\rule[-.02ex]{.25em}{.4pt}\llap{\rule[\ht0]{.25em}{.4pt}}}

\newcommand{\leftouterjoinsymbol}{\mathbin{\ojoin\mkern-5.8mu\bowtie}}

\newcommand{\relalgop}[1]{\textsc{#1}}

\newcommand{\getverticesop}{\iftoggle{textualoperators}{\relalgop{GetVertices}
	}{
		\bigcirc
	}}

\newcommand{\expandbothop}{\iftoggle{textualoperators}{
		\relalgop{ExpandBoth}
	}{
		\updownarrow
	}}

\newcommand{\expandoutop}{\iftoggle{textualoperators}{
		\relalgop{ExpandOut}
	}{
		\uparrow
	}}

\newcommand{\expandinop}{\iftoggle{textualoperators}{
		\relalgop{ExpandIn}
	}{
		\downarrow
	}}

\newcommand{\alldifferentop}{\iftoggle{textualoperators}{
		\relalgop{AllDifferent}
	}{
		\not\equiv
	}}

\newcommand{\duplicateeliminationop}{\iftoggle{textualoperators}{
		\relalgop{DuplicateElimination}
	}{
		\delta
	}}

\newcommand{\sortop}{\iftoggle{textualoperators}{
		\relalgop{Sort}
	}{
		\tau
	}}

\newcommand{\projectionop}{\iftoggle{textualoperators}{
		\relalgop{Projection}
	}{
		\pi
	}}

\newcommand{\selectionop}{\iftoggle{textualoperators}{
		\relalgop{Selection}
	}{
		\sigma
	}}

\newcommand{\groupingop}{\iftoggle{textualoperators}{
		\relalgop{Grouping}
	}{
		\gamma
	}}

\newcommand{\topop}{\iftoggle{textualoperators}{
		\relalgop{Top}
	}{
		\lambda
	}}

\newcommand{\unwindop}{\iftoggle{textualoperators}{
		\relalgop{Unwind}
	}{
		\omega
	}}

\newcommand{\joinop}{\iftoggle{textualoperators}{
		\relalgop{Join}
	}{
		\bowtie
	}}

\newcommand{\leftouterjoinop}{\iftoggle{textualoperators}{
		\relalgop{LeftOuterJoin}
	}{
		\leftouterjoinsymbol
	}}

\newcommand{\unionop}{\iftoggle{textualoperators}{
		\relalgop{Union}
	}{
		\cup
	}}

\newcommand{\bagunionop}{\iftoggle{textualoperators}{
		\relalgop{BagUnion}
	}{
		\uplus
	}}

\newcommand{\intersectionop}{\iftoggle{textualoperators}{
		\relalgop{Intersection}
	}{
		\cap
	}}

\newcommand{\cartesianproductop}{\iftoggle{textualoperators}{
		\relalgop{CartesianProduct}
	}{
		\times
	}}

\newcommand{\getvertices}[2]{\getverticesop_{\vertexvariable{#1}{#2}}}

\newcommand{\kleenestar}{\ast}
\newcommand{\nagivationbody}[3]{\,_{\vertexvariable{#1}{}}^{\vertexvariable{#2}{#3}}}
\newcommand{\expandedgevariable}[4]{
	\left[
	\edgevariable{#1}{#2}
	\ifstrequal{#3}{1} % minHops = 1
	{
		\ifstrequal{#4}{1}
		{} % minHops = 1 and maxHops = 1 -> write nothing
		{\kleenestar_\atom{#3}^\atom{#4}} % minHops = 1 and maxHops != 1
	} % minHops != 1
	{\kleenestar_\atom{#3}^\atom{#4}}
	\right]}

\newcommand{\expandboth}[7]{\expandbothop \nagivationbody{#1}{#2}{#3} \expandedgevariable{#4}{#5}{#6}{#7} }
\newcommand{\expandout}[7]{\expandoutop \nagivationbody{#1}{#2}{#3} \expandedgevariable{#4}{#5}{#6}{#7} }
\newcommand{\expandin}[7]{\expandinop \nagivationbody{#1}{#2}{#3} \expandedgevariable{#4}{#5}{#6}{#7} }

\newcommand{\topp}[2]{\topop_{#1}^{#2}}

\newcommand{\unwind}[1]{\unwindop_{#1}}

\newcommand{\alldifferent}[1]{\alldifferentop_{#1}}
\newcommand{\duplicateelimination}{\duplicateeliminationop}
\newcommand{\sort}[1]{\sortop_{#1}}
\newcommand{\projection}[2]{\projectionop_{#1}^{#2}}

\newcommand{\selection}[1]{\selectionop_{#1}}

\newcommand{\grouping}[2]{\groupingop_{#1}^{#2}}

\newcommand{\join}{\joinop}

\newcommand{\leftouterjoin}{\leftouterjoinop}
\newcommand{\union}{\unionop}
\newcommand{\bagunion}{\bagunionop}

\newcommand{\intersection}{\intersectionop}
\newcommand{\cartesianproduct}{\cartesianproduct}

\definecolor{red}{HTML}{e41a1c}
\definecolor{blue}{HTML}{377eb8}
\definecolor{green}{HTML}{4daf4a}
\definecolor{lilac}{HTML}{984ea3}

\definecolor{progressbargreen}{HTML}{008000}

\newcommand{\externalschemacolorname}{red}
\newcommand{\extravariablescolorname}{blue}
\newcommand{\internalschemacolorname}{green}

\colorlet{externalschemacolor}{\externalschemacolorname}
\colorlet{extravariablescolor}{\extravariablescolorname}
\colorlet{internalschemacolor}{\internalschemacolorname}
\colorlet{nullarynodecolor}{lilac}

\graphicspath{{./figures/}}

\usepackage{amsmath}
\usepackage{mathtools}

\usepackage{chngcntr}
\AtBeginDocument{\counterwithin{lstlisting}{section}}

\renewcommand{\relnull}{\varepsilon}

\newcommand*\numcircledmod[1]{\tikz[baseline=(char.base)]{
		\node[shape=rounded rectangle,rounded corners=1pt,draw,inner sep=1.2pt] (char) {#1};}}

\newcounter{mapping}[section]
\newcommand{\mapping}[1]{\refstepcounter{mapping}\numcircledmod{\themapping}\label{ctr:#1}}
\newcommand{\refmapping}[1]{\numcircledmod{\autoref{ctr:#1}}}

\begin{document}

\newcommand{\gc}{\mathit{gc}}

\newsavebox{\returnbox}
\begin{lrbox}{\returnbox}\lstinline+RETURN+\end{lrbox}
\newcommand{\lstreturn}{\usebox{\returnbox}}

\newsavebox{\withbox}
\begin{lrbox}{\withbox}\lstinline+WITH+\end{lrbox}
\newcommand{\lstwith}{\usebox{\withbox}}

\newsavebox{\matchbox} % ;-)
\begin{lrbox}{\matchbox}\lstinline+MATCH+\end{lrbox}
\newcommand{\lstmatch}{\usebox{\matchbox}}

\newsavebox{\wherebox}
\begin{lrbox}{\wherebox}\lstinline+WHERE+\end{lrbox}
\newcommand{\lstwhere}{\usebox{\wherebox}}

\newsavebox{\distinctbox}
\begin{lrbox}{\distinctbox}\lstinline+DISTINCT+\end{lrbox}
\newcommand{\lstdistinct}{\usebox{\distinctbox}}

\newcommand{\lstreturnwith}{\lstreturn\ / \lstwith\ }

\title{Formalising \opencypher Graph Queries\\ in Relational Algebra}

\titlerunning{Formalising \opencypher Graph Queries in Relational Algebra}

\author{József Marton\inst{1} \and Gábor Szárnyas\inst{2, 3} \and Dániel Varró\inst{2, 3}}

\institute{
	Budapest University of Technology and Economics, Database Laboratory \\
	\email{marton@db.bme.hu} \and
	Budapest University of Technology and Economics \\
	Fault Tolerant Systems Research Group \\
	MTA-BME Lendület Research Group on Cyber-Physical Systems \\
	\email{\{szarnyas, varro\}@mit.bme.hu} \and
	McGill University \\
	Department of Electrical and Computer Engineering
}

\maketitle

\noindent The final publication is available at Springer via \\ \url{https://doi.org/10.1007/978-3-319-66917-5_13}.

\begin{abstract}
	% !TeX spellcheck = en_GB
% !TeX encoding = UTF-8
% The last decade brought considerable improvements in non-relational storage and query technologies, known collectively as NoSQL systems.

Graph database systems are increasingly adapted for storing and processing heterogeneous network-like datasets. However, due to the novelty of such systems, no standard data model or query language has yet emerged.
Consequently, migrating datasets or applications even between related technologies often requires a large amount of manual work or ad-hoc solutions, thus subjecting the users to the possibility of vendor lock-in. To avoid this threat, vendors are working on supporting existing standard languages (\eg SQL) or standardising languages.

In this paper, we present a formal specification for openCypher, a high-level declarative graph query language with an ongoing standardisation effort. We introduce relational graph algebra, which extends relational operators by adapting graph-specific operators and define a mapping from core openCypher constructs to this algebra. We propose an algorithm that allows systematic compilation of openCypher queries.

\end{abstract}

%\tableofcontents

% !TeX spellcheck = en_GB
% !TeX encoding = UTF-8
\section{Introduction}
\label{sec:introduction}

\subsubsection{Context.} Graphs are a well-known formalism, widely used for describing and analysing systems. Graphs provide an intuitive formalism for modelling many real-world scenarios, as the human mind tends to interpret the world in terms of objects (\emph{vertices}) and their respective relationships to one another (\emph{edges})~\cite{CollectivelyGeneratedModel}. 

The \emph{property graph} data model~\cite{DBLP:books/igi/Sakr11/RodriguezN11} extends graphs by adding labels/types and properties for vertices and edges. This gives a rich set of features for users to model their specific domain in a natural way. Graph databases are able to store property graphs and query their contents with complex graph patterns, which otherwise would be are cumbersome to define and/or inefficient to evaluate on traditional relational databases~\cite{TrainBenchmarkSOSYM}.

Neo4j\footnote{\url{https://neo4j.com/}}, a popular NoSQL property graph database, offers the Cypher query language to specify graph patterns. Cypher is a high-level declarative query language which allows the query engine to use sophisticated optimisation techniques. Neo Technology, the company behind Neo4j initiated the \opencypher project~\cite{openCypher},
which aims to deliver an open specification of Cypher.

\subsubsection{Problem and objectives.} The \opencypher project provides a formal specification of the \emph{grammar} of the query language and a set of acceptance tests that define the semantics of various language constructs. This allows other parties to develop their own \opencypher-compatible query engine. However, there is no mathematical formalisation for the language. In ambiguous cases, developers are advised to consult Neo4j's Cypher documentation or to experiment with Neo4j's Cypher query engine and follow its behaviour. Our goal is to provide a formal specification for the core features of \opencypher.

\subsubsection{Contributions.} In this paper, we use a formal definition of the property graph data model~\cite{DBLP:conf/edbt/HolschG16} and relational graph algebra, which operates on multisets (bags)~\cite{DBLP:books/daglib/0020812} and is extended with additional graph-specific operators. Using these foundations, we construct a concise formal specification for the core features in the \opencypher grammar. This specification is detailed enough to serve as a basis for an \opencypher compiler~\cite{openCypherReport}.

% !TeX spellcheck = en_GB
% !TeX encoding = UTF-8
\section{Data Model and Running Example}

\subsubsection{Data model.}
%\label{sec:property-graph}

A \emph{property graph} is defined as $G = (V, E, \verticestoedges, \vertexlabels, \edgelabels, \vertexlabelfunction, \edgelabelfunction, \vertexproperties, \edgeproperties)$, where $V$ is a set of vertices, $E$ is a set of edges and $\verticestoedges: E \assign V \cartesianproductop V$ assigns the source and target vertices to edges. Vertices are labelled and edges are typed:
\begin{itemize}
	\item $\vertexlabels$ is a set of vertex labels, $\vertexlabelfunction: V \assign 2^{\vertexlabels}$ assigns a \emph{set of labels} to each vertex.
	\item $\edgelabels$ is a set of edge types, $\edgelabelfunction: E \assign \edgelabels$ assigns a \emph{single type} to each edge.
\end{itemize}

To define properties, let $D = \cup_{i} D_i$ be the union of atomic domains $D_i$ and let $\relnull$ represent the \textsf{NULL} value.
\begin{itemize}
	\item $\vertexproperties$ is a set of vertex properties. A vertex property $p_i \in \vertexproperties$ is a partial function $p_i: V \assign D_i \unionop \{ \relnull \}$, which assigns a property value from a domain $D_i \in D$ to a vertex $v \in V$, if $v$ has property $p_i$, otherwise $p_i(v)$ returns $\relnull$.
	\item $\edgeproperties$ is a set of edge properties. An edge property $p_j \in \edgeproperties$ is a partial function $p_j: E \assign D_j \unionop \{ \relnull \}$, which assigns a property value from a domain $D_j \in D$ to an edge $e \in E$, if $e$ has property $p_j$, otherwise $p_j(e)$ returns $\relnull$.
\end{itemize}

In the context of this paper, we define a \emph{relation} as a \emph{bag} (\emph{multiset}) of tuples: a tuple can occur more than once in a relation~\cite{DBLP:books/daglib/0020812}.
Given a property graph $G$, relation $r$ is a \emph{graph relation} if the following holds:
$$\forall A \in \schema{r}: \dom{A} \subseteq V \union E \union D,$$
where the schema of $r$, $\schema{r}$, is a list containing the attribute names, $\dom{A}$ is the domain of attribute $A$, $V$ is the vertices of $G$, and $E$ is the edges of $G$.

\subsubsection{Property access.} When defining relational algebra expression on graph relations, it is often required (\eg in projection and selection operators) to access a certain property of a vertex/edge. Following the notation of~\cite{DBLP:conf/edbt/HolschG16}, if $x$ is an attribute of a graph relation, we use $x.p$ to access the corresponding value of property $p$. Also, $\vertexlabelfunction(v)$ returns the labels of vertex $v$ and $\edgelabelfunction(e)$ returns the type of edge $e$.

\subsubsection{Running example.}
\label{sec:running-example}

\begin{figure}[t]
	\centering
    \begin{subfigure}[b]{0.5\textwidth}
		\centering
		\includegraphics[width=\textwidth]{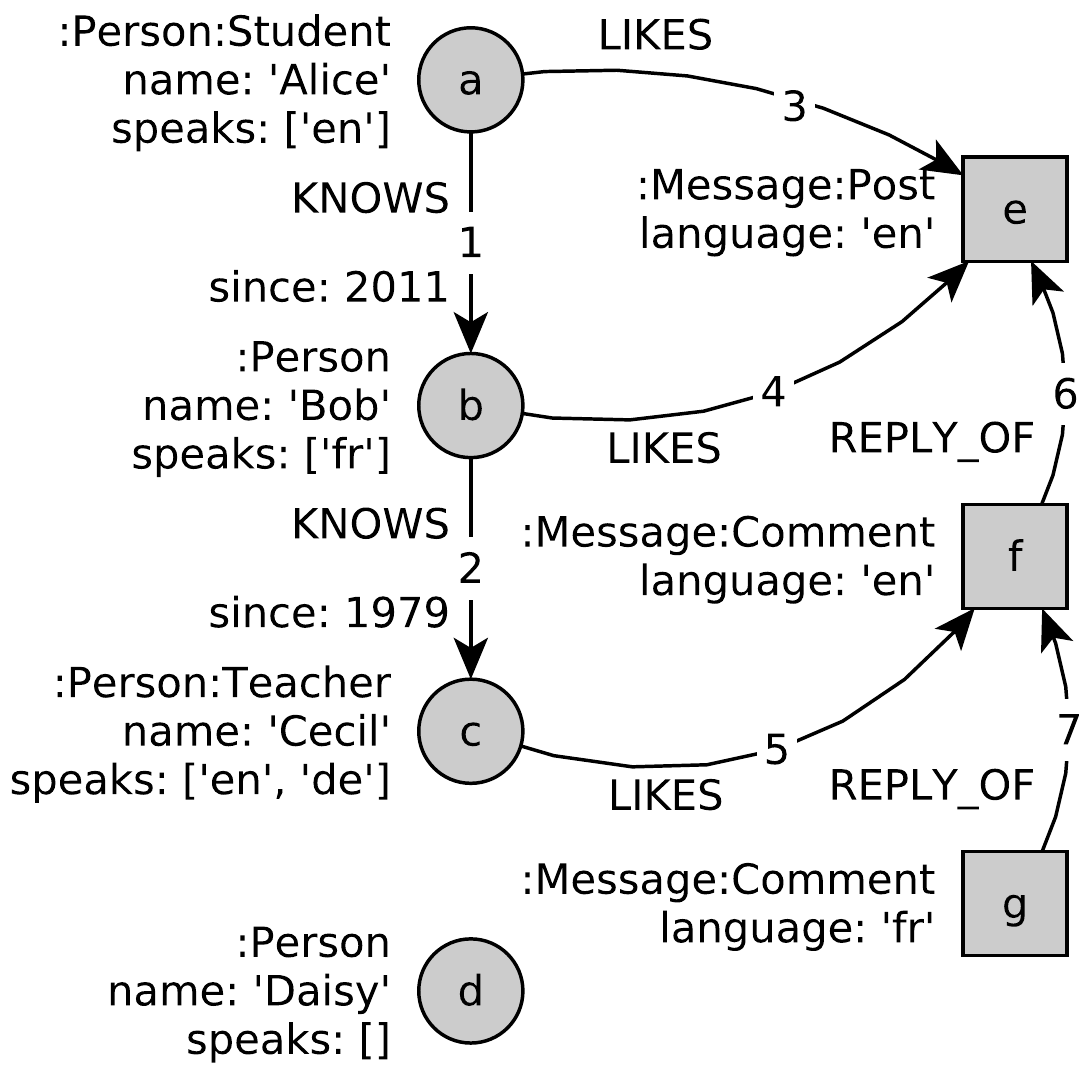}
		\caption{Example social network graph.}
		\label{fig:running-example-graph}
    \end{subfigure}%
	~
	\begin{subfigure}[b]{0.5\textwidth}
		\input{running-example-formalised}
		\caption{The dataset as a property graph.}
		\label{fig:running-example-formalised}
	\end{subfigure}
	\caption{Social network example represented graphically and formally. To improve readability, we use letters for vertex identifiers and numbers for edge identifiers.}
\end{figure}

An example graph inspired by the LDBC Social Network Benchmark~\cite{DBLP:conf/sigmod/ErlingALCGPPB15} is shown on \autoref{fig:running-example-graph}, while \autoref{fig:running-example-formalised} presents the formalised graph. The graph vertices model four \textsf{Persons} and three \textsf{Messages}, with edges representing \textsf{LIKES}, \textsf{REPLY\_OF} and \textsf{KNOWS} relations. In social networks, the \textsf{KNOWS} relation is symmetric, however, the property graph data model does not allow undirected edges. Hence, we use directed edges with an arbitrary direction and model the symmetric semantics of the relation in the queries.

% !TeX spellcheck = en_GB
% !TeX encoding = UTF-8
\section{The openCypher Query Language}
\label{sec:opencypher}

Cypher is the a high-level declarative graph query language of the Neo4j graph database. It allows users to specify graph patterns with a syntax resembling an actual graph, which makes the queries easy to comprehend. The goal of the \opencypher project~\cite{openCypher} is to provide a standardised specification of the Cypher language.
In the following, we introduce features of the language using examples.

\subsection{Language Constructs}
\label{sec:language-constructs}

\subsubsection{Inputs and outputs.} \opencypher queries take a \emph{property graph} as their input, however the result of a query is not a graph, but a \emph{graph relation}.

\subsubsection{Vertex and path patterns.}

The basic building blocks of queries are patterns of vertices and edges.
\autoref{lst:get-vertices} shows a query that returns all vertices that model a \textsf{Person}. The query in \autoref{lst:expand-out} matches \textsf{Person} and \textsf{Message} pairs connected by a \textsf{LIKES} edge and returns the person's name and the message language. \autoref{lst:multihop} describes person pairs that know each other directly or have a friend in common, \ie from person $\atom{p1}$, the other person $\atom{p2}$ can be reached using one or two hops.

\vspace{-3ex}

\begin{figure}[H]
  \centering
  \begin{minipage}{.35\textwidth}
    \centering
    \listingcypher{get-vertices}{Getting vertices}
  \end{minipage}
  \qquad
  \begin{minipage}{.55\textwidth}
    \centering
    \listingcypher{expand-out}{Pattern matching}
  \end{minipage}
\end{figure}

\vspace{-6.3ex}

\begin{figure}[H]
        \centering
        \begin{minipage}{.45\textwidth}
	        \centering
        	\listingcypher{multihop}{Variable length path}
        \end{minipage}
        \qquad
        \begin{minipage}{.45\textwidth}
	        \centering
                \listingcypher{grouping}{Grouping}
        \end{minipage}
\end{figure}

\subsubsection{Filtering.}

Pattern matches can be filtered in two ways as illustrated in \autoref{lst:selection1} and \autoref{lst:selection2}. (1) Vertex and edge patterns in the \lstinline+MATCH+ clause might have vertex label/edge type constraints written in the pattern after a colon, and (2) the optional \lstinline+WHERE+ subclause of \lstinline+MATCH+ might hold predicates.

\vspace{-1ex}

\begin{figure}[H]
  \centering
  \begin{minipage}{.55\textwidth}
    \centering
    \listingcypher{selection1}{Filtering for edge property}
  \end{minipage}
  \qquad
  \begin{minipage}{.35\textwidth}
    \centering
    \listingcypher{selection2}{Filtering}
  \end{minipage}
\end{figure}

\subsubsection{Unique and non-unique edges.}
\label{sec:oc-uniqueness}

A \lstinline+MATCH+ clause defines a graph pattern. A query can be composed of multiple patterns spanning multiple \lstinline+MATCH+ clauses. For matches of a pattern within a single \lstinline+MATCH+ clause, edges are required to be unique.
However, matches for multiple \lstinline+MATCH+ clauses can share edges.
This means that in matches returned by \autoref{lst:alldifferent}, $\atom{k1}$ and $\atom{k2}$ are required to be different, while in matches returned by \autoref{lst:non-unique-edges}, $\atom{k1}$ and $\atom{k2}$ are allowed to be equal.
For vertices, this restriction does not apply.\footnote{Requiring uniqueness of edges is called \emph{edge isomorphic matching}. Other query languages and execution engines might use \emph{vertex isomorphic matching} (requiring uniqueness of vertices), \emph{isomorphic matching} (requiring uniqueness of both vertices and edges) or \emph{homomorphic matching} (not requiring uniqueness of either)~\cite{DBLP:conf/sigmod/Junghanns17}.} 
This is illustrated in \autoref{lst:triangle}, which returns adjacent persons who like the same message.

\vspace{-2.5ex}

\begin{figure}[H]
  \centering
  \begin{minipage}{.45\textwidth}
    \centering
    \listingcypher{alldifferent}{Different edges}
  \end{minipage}
  \qquad
  \begin{minipage}{.45\textwidth}
    \centering
    \listingcypher{non-unique-edges}{Non-unique edges}
  \end{minipage}
\end{figure}

\listingcypher{triangle}{Triangle}

\subsubsection{Creating the result set.}

The result set\footnote{The term \emph{result set} refers to the \emph{result collection}, which can be a set, a bag or a list.}
of a query is basically given in the \lstinline+RETURN+ clause, 
which can be de-duplicated using the \lstinline+DISTINCT+ modifier, sorted using the \lstinline+ORDER BY+ subclause. Skipping rows after sorting and limiting the result set to a certain number of records can be achieved using \lstinline+SKIP+ and \lstinline+LIMIT+ modifiers.

\autoref{lst:duplicate-elimination-sorting} illustrates these concepts by returning the name of the persons. The result set is restricted to the second and third names in alphabetical order.

\vspace{-2.5ex}

\begin{figure}[H]
  \centering
  \begin{minipage}{.4\textwidth}
    \centering
    \listingcypher{duplicate-elimination-sorting}{Deduplicate and sort}
  \end{minipage}
  \quad\enskip
  \begin{minipage}{.54\textwidth}
    \centering
    \lstset{numbers=left}
    \listingcypher{join}{Multiple patterns}
    \lstset{numbers=none}
  \end{minipage}
\end{figure}

\subsubsection{Combining patterns.}

Multiple patterns (in the same or in different) \lstinline+MATCH+ clauses are combined together based on their common variables. \autoref{lst:join} illustrates this by showing two patterns on lines 2 and 3. The first pattern describes a message $\atom{m}$ that has at least two likes.
The second pattern finds replies to $\atom{m}$.

\vspace{-3ex}

\subsubsection{Aggregation.}
\label{sec:aggregation}

\opencypher specifies aggregation operators for performing calculations on multiple tuples.\footnote{The \mylstc{avg} \mylstc{count} \mylstc{max} \mylstc{min} \mylstc{percentileCont} \mylstc{percentileDisc} \mylstc{stdDev} \mylstc{stdDevP} \mylst{sum} functions return a single scalar value, while \mylst{collect} returns a list.} Unlike in SQL queries,
the \emph{grouping criteria} is determined implicitly in the \mylst{RETURN} as well as in 
and \mylst{WITH} clauses. Each expression of the expression list in \mylst{WITH} and \mylst{RETURN} are forced to contain either (1)~no aggregate functions or (2)~a single aggregate function at the outermost level. The grouping key is the tuple built from expressions of type~(1).\footnote{Decision on grouping semantics is due after the camera ready submission deadline. The semantics presented in this paper is one of the possible approaches.}
The query of \autoref{lst:grouping} counts the number of persons commanding each language.

\vspace{-3.5ex}

\begin{figure}[H]
	\centering
	\begin{minipage}[b]{.36\textwidth}
		\centering
		\listingcypher{unwind}{Unwind}
	\end{minipage}
	\qquad
	\begin{minipage}[b]{.5\textwidth}
		\centering
		\scriptsize
		\begin{tabular}{|l|c|}
			\hline
			\bf p.name & \bf lang \\ \hline\hline
			Alice & en \\ \hline
			Bob & fr \\ \hline
			Cecil & en \\ \hline
			Cecil & de \\ \hline
		\end{tabular}
		\vspace{-2ex}
		\caption{Output of the unwind query.}
		\label{fig:unwind}
		\vspace{-5ex}
	\end{minipage}
\end{figure}

\subsubsection{Unwinding a list.}

The \lstinline+UNWIND+ construct takes an attribute and multiplies each tuple by appending the list elements one by one to the tuple, thus modifying the schema of the query part. By applying \lstinline+UNWIND+ to the $\atom{speaks}$ attribute \autoref{lst:unwind} lists persons along the languages they speak. \autoref{fig:unwind} shows the output of this query. As Cecil speaks two languages, he appears twice in the output. Note that ``Daisy'' speaks no languages, thus no tuples belong to her in the output.

\subsection{Query Structure}

In \opencypher a query is composed as the \lstinline+UNION+ of one or more single queries. Each single query must have the same resulting schema, i.e. the resulting tuples must have the same arity and the same name at each position.

\subsubsection{Single queries.} A single query is composed of one or more query parts written subsequently. Query parts that form a prefix of a single query have one result set with the schema of the last query part's schema in that prefix.

\subsubsection{Query parts.} Clause sequence of a query part matches the regular expression as follows: \mbox{\lstinline+MATCH*((WITH UNWIND?)|UNWIND|RETURN)+}. They begin with an arbitrary number of \lstinline+MATCH+ clauses, followed by either (1)~\lstinline+WITH+ and an optional \lstinline+UNWIND+, (2)~a single \lstinline+UNWIND+, or (3)~a \lstinline+RETURN+ in case of the last query part.\footnote{In \opencypher, the filtering \lstinline+WHERE+ operation is a subclause of \lstinline+MATCH+ and \lstinline+WITH+. When used in \lstinline+WITH+ as illustrated on line 3 of \autoref{lst:multiple-subqueries}, \lstinline+WHERE+ is similar to the \lstinline+HAVING+ construct of SQL with the major difference that, in \opencypher it is also allowed when no aggregation was specified in the query.}

The \lstinline+RETURN+ and \lstinline+WITH+ clauses use similar syntax and have the same semantics, the only difference being that \lstinline+RETURN+ should be used in the last query part while \lstinline+WITH+ should only appear in the preceding ones. These clauses list expressions whose value form the tuples, thus they determine the schema of the query parts.

\vspace{-2.5ex}

\begin{figure}[H]
	\centering
	\begin{minipage}[c]{.7\textwidth}
		\centering
		\lstset{numbers=left}
		\listingcypher{multiple-subqueries}{Single query with multiple query parts}
		\lstset{numbers=none}
	\end{minipage}
	\quad
	\begin{minipage}[c]{.25\textwidth}
		\centering
		\begin{tabular}{|c|c|}
			\hline
			\bf reply & \bf orig \\ \hline\hline
			fr & en \\ \hline
		\end{tabular}
		\vspace{-2ex}
		\caption{Result.}
		\label{fig:multiple-subqueries-result}
	\end{minipage}
\end{figure}

%\vspace{-0.5ex}

\subsubsection{Example.} An \opencypher single query composed of two query parts is shown on \autoref{lst:multiple-subqueries} along with its result on \autoref{fig:multiple-subqueries-result}. It retrieves the language of messages that were written in a language no other message uses. If that message was a reply, the language of the original message is also retrieved.

The first query part spans lines 1--3 and the second spans lines 4--6. The result of the first query part is a single tuple $\tuple{\atom{``fr"}, 1}$ with the schema $\tuple{\atom{singleLang}, \atom{cnt}}$. The second query part takes this result as an input to retrieve messages of the given languages and in case of a reply the original message in $\atom{m3}$. 
The result of these two query parts together produces the final result whose schema is determined by the \lstinline+RETURN+ of the last query part (line 6).

% !TeX spellcheck = en_GB
% !TeX encoding = UTF-8
\section{Mapping \opencypher to \RGA}

In this section, we present relational graph algebra using the examples of \autoref{sec:language-constructs} and provide a mapping that allows compilation from \opencypher to this algebra.

\newcommand{\propheader}{\multirow{1}{*}{\bf props.}}

\setlength\tabcolsep{3.6pt}
\begin{table}[htbp]
	\centering
	\begin{tabular}{|c|c|c|c|c|}
		\hline
		\multirow{1}{*}{\bf \#ops.} &             \multirow{1}{*}{\bf notation}              & \multirow{1}{*}{\bf name} & \propheader &                   \multirow{1}{*}{\bf schema}                    \\ \hline\hline
		 \multirow{1}{*}{\bf 0}   &                  $\getvertices{v}{L}$                   &     \getverticestext      &     $-$     &                        $\tuple{\var{v}}$                        \\ \hline\hline
		 \multirow{9}{*}{\bf 1}   &  $\expandboth{v}{w}{L}{e}{T}{1}{1} (r)$  &      \expandbothtext      &     $-$     &             $\schema{r} \append \tuple{\var{e}, \var{w}}$             \\ \cline{2-5}
		                          &         $\alldifferent{\atom{variables}} (r)$          &     \alldifferenttext     &      i      &                           $\schema{r}$                           \\ \cline{2-5}
		                          &              $\unwind{\var{xs} \assign \var{x}} (r)$               &        \unwindtext        &     $-$     & $\schema{r} \remove \tuple{\var{xs}} \append \tuple{\var{x}} $ \\ \cline{2-5}
		            %             &               $\nest{xs \assign x} (r)$                &         \nesttext         &     $-$     & $\schema{r} \remove \tuple{\atom{x}} \append \tuple{\atom{xs}} $ \\ \cline{2-5}
		            %             &              $\unnest{xs \assign x} (r)$               &        \unnesttext        &     $-$     & $\schema{r} \remove \tuple{\atom{xs}} \append \tuple{\atom{x}} $ \\ \hline\hline
		                          &           $\selection{\atom{condition}} (r)$           &      \selectiontext       &      i      &                           $\schema{r}$                           \\ \cline{2-5}
		                          &  $\projection{\var{x_1}, \var{x_2}, \ldots}{} (r)$   &      \projectiontext      &      i      &                $\tuple{\var{x_1}, \var{x_2}, \ldots}$                 \\ \cline{2-5}
		                          &   $\grouping{\var{x_1}, \var{x_2}, \ldots}{\var{c_1}, \var{c_2}, \ldots} (r)$    &       \groupingtext       &      i      &                $\tuple{\var{x_1}, \var{x_2}, \ldots}$                 \\ \cline{2-5}
		                          &             $\duplicateeliminationop (r)$              & \duplicateeliminationtext &      i      &                           $\schema{r}$                           \\ \cline{2-5}
		                          & $\sort{\desc \var{x_1}, \asc \var{x_2}, \ldots} (r)$ &         \sorttext         &      i      &                           $\schema{r}$                           \\ \cline{2-5}
		                          &         $\topp{\atom{skip}}{\atom{limit}} (r)$         &         \toptext          &     $-$     &                           $\schema{r}$                           \\ \hline\hline
		 \multirow{4}{*}{\bf 2}   &                     $r \unionop s$                     &        \uniontext         &     $-$     &                           $\schema{r}$                           \\ \cline{2-5}
		                          &                    $r \bagunion s$                     &       \baguniontext       &    c, a     &                           $\schema{r}$                           \\ \cline{2-5}
		                          &                     $r \joinop s$                      &         \jointext         &    c, a     &      $\schema{r} \append (\schema{s} \remove \schema{r}) $       \\ \cline{2-5}
		                          &                  $r \leftouterjoin s$                  &    \leftouterjointext     &     $-$     &      $\schema{r} \append (\schema{s} \remove \schema{r}) $       \\ \cline{1-5}
		                          %&                   $r \antijoinop s$                    &       \antext       &    c, a     &                           $\schema{r}$                           \\ \cline{1-5}
	\end{tabular}
	\caption{Number of operands, properties and result schemas of relational graph algebra operators. A unary operator $\alpha$ is idempotent~(i), iff $\alpha(x) = \alpha(\alpha(x))$ for all inputs. A binary operator $\beta$ is commutative~(c), iff $x~\beta~y = y~\beta~x$ and associative~(a), iff $(x~\beta~y)~\beta~z = x~\beta~(y~\beta~z)$. For schema transformations, append is denoted by $\append$, while removal is marked by $\remove$. $L$ represents a (possibly empty) set of vertex labels and $T$ represents a (possibly empty) set of edge types.}
\label{table:operators}
\end{table}

\subsection{An Algebra for Formalising Graph Queries}
\label{sec:rga}

We present both standard operators of relational algebra~\cite{DBLP:books/daglib/0006733} and operators for graph relations. \autoref{table:operators} provides an overview of the operators of \rga. We follow the \opencypher query language and present a mapping from the language constructs to their algebraic equivalents\footnote{Patterns in the \opencypher query might contain anonymous vertices and edges. In the algebraic form, we denote this with names starting with an underscore, such as $\var{\_v1}$ and $\var{\_e2}$.}, summarized in \autoref{table:mapping}. The corresponding rows of the table (\eg \refmapping{getvertices}) are referred to in the text.

%%%%%%%%%%%%%%%%%%%%%%%%%%%%%%%%%%%%%%%%%%%%%%%%%%%%%%%%%%%%%%%%%%%%%%%%%%%%%%%%

\subsubsection{Basic operators.}

The \emph{\projectiontext} operator $\projectionop$ keeps a specific set of attributes in the relation: $ t = \projection{\var{x_1}, \ldots, \var{x_n}}{} \left(r\right).$ Note that the tuples are not deduplicated (\ie filtered to sets), thus $t$ will have the same number of tuples as $r$. The projection operator can also rename the attributes, \eg $\projection{\var{x1} \assign \var{y1}}{} \left(r\right)$ renames $\var{x1}$ to $\var{y1}$.

The \emph{\selectiontext} operator $\selectionop$ filters the incoming relation according to some criteria. Formally,
$ t = \selection{\theta} \left(r\right), $
where predicate $\theta$ is a propositional formula. Relation $t$ contains all tuples from $r$ for which $\theta$ holds.

\subsubsection{Vertices and patterns.} \refmapping{getvertices}--\refmapping{getverticeslabels} The \emph{\getverticestext}~\cite{DBLP:conf/edbt/HolschG16} nullary operator $\getvertices{v}{l_1 \land \ldots \land l_n}$ returns a graph relation of a single attribute $v$ that contains vertices that have \emph{all} of labels $l_1, \ldots, l_n$. Using this operator, the query in \autoref{lst:get-vertices} is compiled to
$$
%\projection{\var{p}
%}{}
\getvertices{p}{Person}
$$

\refmapping{patterni}--\refmapping{patterniv} The \emph{\expandouttext} unary operator $\expandout{v}{w}{l_1 \land \ldots \land l_n}{e}{t_1 \lor \ldots \lor t_k}{1}{1}(r)$ adds new attributes $e$ and $w$ to each tuple iff there is an edge $e$ from $v$ to $w$, where
$e$ has \emph{any} of types $t_1, \ldots, t_k$, while
$w$ has \emph{all} labels $l_1, \ldots, l_n$.\footnote{Label and type constraints can be omitted for the \getverticestext operator and the expand operators. For example, $\getvertices{v}{}$ returns all vertices, while $\expandout{v}{w}{}{e}{}{1}{1} (r)$ traverses all outgoing edges $e$ from vertices $v$ to $w$, regardless of their labels/types.}
More formally, this operator appends the $\tuple{e, w}$ to a tuple iff $\verticestoedges(e) = \tuple{v, w}$, $l_1, \ldots, l_n \in \mathcal{L}(w)$ and $\mathcal{T}(e) \in \{t_1, \ldots, t_k\}$. Using this operator, the query in \autoref{lst:expand-out} can be formalised as
$$
\projection{\var{p.name}, \var{m.language}
}{}
\expandout{p}{m}{Message}{\_e1}{LIKES}{1}{1}
\getvertices{p}{Person}
$$

Similarly to the \expandouttext operator, the \emph{\expandintext} operator~$\expandinop$ appends $\tuple{e, w}$ iff $\verticestoedges(e) = \tuple{w, v}$, while the \emph{\expandbothtext} operator~$\expandbothop$ uses edge $e$ iff either $\verticestoedges(e) = \tuple{v, w}$ or $\verticestoedges(e) = \tuple{w, v}$.
We also propose an extended version of this operator, $\expandout{v}{w}{}{e}{}{min}{max}$, which may use between $\atom{min}$ and $\atom{max}$ hops. % With the default setting, \ie $\atom{min}=\atom{max}=1$, a single edge variable $e$ can be used instead of edge list $E$.
Using this extension, \autoref{lst:multihop} is compiled to
$$
\projection{\var{p1}, \var{p2}
}{}
%\Big(
\alldifferent{\var{ks}}\,\,%\Big(
\expandboth{p1}{p2}{Person}{ks}{KNOWS}{1}{2}%\Big(
\getvertices{p1}{Person}
%\Big)
%\Big)
%\Big)
$$

%%%%%%%%%%%%%%%%%%%%%%%%%%%%%%%%%%%%%%%%%%%%%%%%%%%%%%%%%%%%%%%%%%%%%%%%%%%%%%%%

\subsubsection{Combining and filtering pattern matches.}

%\paragraph{Uniqueness of edges.}
\refmapping{matchi}--\refmapping{optionalmatchii} In order to express the uniqueness criterion for edges (illustrated in \autoref{sec:oc-uniqueness}) in a compact way, we propose the \emph{\alldifferenttext} operator. The \alldifferenttext operator $\alldifferent{E_1, \ldots, E_n}{(r)}$ filters $r$ to keep tuples where variables in $\cup_{i} E_{i}$ are pairwise different.
Note that the operator is actually a shorthand for the following selection:
$$\alldifferent{E_1, \ldots, E_n}{(r)} = \selection{ \hspace{-3.2ex} \bigwedge\limits_{\substack{e_1, e_2 \in \cup_{i} {E_i} \\ e_1 \neq e_2}} { \hspace{-3.8ex} r.e_1 \,\neq\, r.e_2 } }{(r)}$$

\noindent Using the \alldifferenttext operator, query in \autoref{lst:alldifferent} is compiled to
$$
\projection{\var{p1}, \var{k1}, \var{p2}, \var{k2}, \var{p3}
}{}
\alldifferent{\var{k1}, \var{k2}}
\expandout{p2}{p3}{}{k2}{KNOWS}{1}{1}
\expandout{p1}{p2}{}{k1}{KNOWS}{1}{1}
\getvertices{p1}{}
$$

\refmapping{matchi}--\refmapping{matchii} %The $\cartesianproductop$ operator produces the \emph{\cartesianproducttext} $ t = r \cartesianproductop s.$
The result of the \emph{\jointext} operator $\joinop$ is determined by creating the \cartesianproducttext of the relations, then filtering those tuples which are equal on the attributes that share a common name. The combined tuples are projected: from the attributes present in both of the two input relations, we only keep the ones in $r$ and drop the ones in $s$. Thus, the join operator is defined as
$$r \join s = \pi_{R \union S} \left(\selection{r.A_1 = s.A_1\,\land\,\ldots\,\land\,r.A_n = s.A_n)} \left(r \times s\right) \right),$$
where $ \{ A_1, \ldots, A_n \} = R \intersection S $ is the set of attributes that occur both in $R$ and $S$. % Note that if the set of common attributes is empty, the \jointext operator is equivalent to the Cartesian product of the relations.

\noindent In order to allow pattern matches to share the same edge, they must be included in different \lstinline+MATCH+ clauses as shown on \autoref{lst:non-unique-edges} which is compiled to
$$
\projection{\var{p1}, \var{p2}, \var{p3}
}{}
\bigg(
\Big(
\expandboth{p1}{p2}{}{\_e1}{KNOWS}{1}{1}
\getvertices{p1}{}
\Big)
\join
\,
\Big(
\expandboth{p2}{p3}{}{\_e2}{KNOWS}{1}{1}
\getvertices{p2}{}
\Big)
\bigg)
$$

The query in \autoref{lst:join} with two patterns in one \lstinline|MATCH| clause is compiled to:
\begin{align*}
\begin{autobreak}
\projection{\var{r}
}{}
%\Big(
\alldifferent{\var{\_e1}, \var{\_e2}, \var{\_e3}}\bigg(
\Big(
\expandin{m}{\_v2}{}{\_e2}{LIKES}{1}{1}
\expandout{\_v1}{m}{Message}{\_e1}{LIKES}{1}{1}%\Big(
\getvertices{\_v1}{}
%\Big)
\Big)
\quad
\join
\Big(
\expandin{m}{r}{}{\_e3}{REPLY\_OF}{1}{1}%\Big(
\getvertices{m}{Message}
\Big)
\bigg)
%\Big)
\end{autobreak}
\end{align*}

%\input{relalg-triangle}

%The \leftouterjointext $\myleftouterjoin$ pads tuples from the left relation that did not match any from the right relation with $\relnull$ values and adds them to the result of the \jointext~\cite{DBLP:books/daglib/0015084}.
\refmapping{optionalmatchi}--\refmapping{optionalmatchii} The \emph{\leftouterjointext} operator produces $t = r \leftouterjoinop s$ combining matching tuples of $r$ and $s$ according to a given matching semantics.\footnote{Matching semantics might use value equality of attributes that share a common name (similarly to \jointext) to use an arbitrary condition (similarly to $\theta$-join).} In case there is no matching tuple in $s$ for a particular tuple $e \in r$, $e$ is still included in the result, with tuple attributes that exclusively belong to relation $s$ having a value of $\relnull$.

%%%%%%%%%%%%%%%%%%%%%%%%%%%%%%%%%%%%%%%%%%%%%%%%%%%%%%%%%%%%%%%%%%%%%%%%%%%%%%%%

\subsubsection{Result and subresult operations.}

\refmapping{returnii} The \emph{\duplicateeliminationtext} operator $\duplicateeliminationop$ eliminates duplicate tuples in a bag.

\refmapping{returniii} The \emph{\groupingtext} operator $\groupingop$ groups tuples according to their value in one or more attributes and aggregates the remaining attributes. %Aggregated values (scalars and inline collections) makes \rga not closed under \groupingtext.

We generalize the \groupingtext operator to explicitly state the \emph{grouping criteria} and allow for complex aggregate expressions.
This is similar to the SQL query language where the grouping criteria is explicitly given in \mylst{GROUP BY}.

We use the notation $\grouping{e_1, e_2, \ldots }{c_1, c_2, \ldots }$, where $c_1, c_2, \ldots $ in the superscript form the \emph{grouping criteria}, \ie the list of expressions whose values partition the incoming tuples into groups. For each and every group this aggregation operator emits a single tuple of expressions listed in the subscript, \ie $\tuple{e_1, e_2, \ldots }$.
Given attributes $\{a_1, \ldots, a_n\}$ of the input relation, $c_i$ is an arithmetic expression built from $a_j$ attributes using common arithmetic operators, while $e_i$ is an expression built from $a_j$ using common arithmetic operators and grouping functions. % like in~\cite{DBLP:books/daglib/0020812}: \lstinline+SUM+, \lstinline+AVG+, \lstinline+MIN+, \lstinline+MAX+ and \lstinline+COUNT+.

We have discussed the aggregation semantics of \opencypher in \autoref{sec:aggregation}. The formal algorithm for determining the grouping criteria is given in \autoref{alg:grouping-criteria}.
Building on this algorithm and the \groupingtext operator, \autoref{lst:grouping} is compiled to
$$
\grouping{\var{language}, \literal{count\_distinct}
	\left( \var{p.name} \right)
	\assign \var{cnt}}{\var{language}}
\unwind{\var{p.speaks} \assign \var{language}}
\getvertices{p}{Person}
$$

\vspace{-5ex}

\SetKwFunction{KwFnDGC}{DetermineGroupingCriteria}
\begin{algorithm}[htb]
  \DontPrintSemicolon
  \KwData{E is the list of expressions in the \lstreturn\ or \lstwith\ clause}
  \Fn{\KwFnDGC{E}}{
    $G \leftarrow \{\}$ \tcp*[r]{initial set of grouping criteria}
    \ForEach{$e \in E$} {
      \uIf{e has an aggregate function call at its outermost level}{
        \tcp*[r]{do nothing as this is an aggregation}
      }
      \uElseIf{e contains aggregate function call}{
        \tcp*[r]{aggregation allowed only at the outermost level}
        {\bf raise} SemanticError(Illegal~use~of~aggregation~function) \;
      }
      \Else{
        $G \leftarrow G \union \{e\}$ \tcp*[r]{append to the grouping key}
      }
    }
    \Return $G$
  }
  \caption{Determine grouping criteria from return item list.}
  \label{alg:grouping-criteria}
\end{algorithm}
%
%\SetKwFunction{KwFnDGC}{DetermineGroupingCriteria}
%\begin{algorithm}[htb]
%  \DontPrintSemicolon
%  \KwData{E is the list of elements in the \lstreturn\ or \lstwith\ clause}
%  \Fn{\KwFnDGC{E}}{
%    $G \leftarrow \{\}$ \tcp*[r]{initial set of grouping criteria}
%    $Q \leftarrow E$ \tcp*[r]{queue of expressions to process}
%    \While{Q is not empty} {
%      $q \leftarrow \mathrm{pop}(Q)$\;
%      \uIf(\tcp*[f]{vertex, edge or property}){q is a variable}{
%        $G \leftarrow G \union \{q\}$ \;
%      }
%      \uElseIf{q is a non-grouping function}{
%        $Q \leftarrow Q \union \mathrm{arguments}\left(q\right)$ \tcp*[r]{add each argument}
%      }
%      \ElseIf(\tcp*[f]{list, map, etc.}){q is a compound expression}{
%        $Q \leftarrow Q \union \{q\}$ \tcp*[r]{add each element}
%      }
%    }
%    \Return $G$
%  }
%  \caption{Determine grouping criteria from return item list.}
%  \label{alg:grouping-criteria}
%\end{algorithm}

%\vspace{-3ex}

%%%%%%%%%%%%%%%%%%%%%%%%%%%%%%%%%%%%%%%%%%%%%%%%%%%%%%%%%%%%%%%%%%%%%%%%%%%%%%%%\

\subsubsection{Unwinding and list operations.}

%Most textbooks also define \emph{extended operators} of relational algebra~\cite{DBLP:books/daglib/0020812}:

\refmapping{unwind} The \emph{\unwindtext}~\cite{DBLP:conf/dlog/BotoevaCCRX16} operator $\unwind{\var{xs} \assign \var{x}}$ takes the list in attribute $\var{xs}$ and multiplies each tuple adding the list elements one by one to an attribute $\var{x}$, as demonstrated in \autoref{fig:unwind}. Using this operator, the query in \autoref{lst:unwind} can be formalised as
$$
\projection{\var{p.name}, \var{lang}
}{}
\unwind{\var{p.speaks} \assign \var{lang}}
\projection{\var{p}
}{}
\getvertices{p}{Person}
$$

\refmapping{orderby} The \emph{\sorttext} operator $\sortop$ transforms a bag relation of tuples to a list of tuples by ordering them. The ordering is defined for selected attributes and with a certain direction for each of them (ascending $\asc$/descending $\desc$), \eg $\sortop_{\asc \var{x1}, \desc \var{x2}} (r)$.

\refmapping{top} The \emph{\toptext} operator $\topp{l}{s}$ (adapted from~\cite{DBLP:conf/sigmod/LiCIS05}) takes a list as its input, skips the first $s$ tuples and returns the next $l$ tuples.\footnote{SQL implementations offer the \texttt{OFFSET} and the \texttt{LIMIT}/\texttt{TOP} keywords.}

Using the \sorttext and \toptext operators, the query of \autoref{lst:duplicate-elimination-sorting} is compiled to:
$$
\topp{2}{1}
\sort{\asc \var{p.name}}
\duplicateelimination
\,
\projection{\var{p.name}
}{}
\getvertices{p}{Person}
$$

\subsubsection{Combining results.}

The $\unionop$ operator produces the \emph{set union} of two relations, while the $\bagunionop$ operator produces the \emph{\baguniontext} of two operators, \eg $\{\tuple{1, 2}, \tuple{3, 4}\} \bagunionop \{\tuple{1, 2}\} = \{\tuple{1, 2}, \tuple{1, 2}, \tuple{3, 4}\}$. For both the \uniontext and \baguniontext operators, the schema of the operands must have the same attributes.

\subsection{Mapping \opencypher Queries to \RGA}
\label{sec:compilation}

In this section, we give the mapping algorithm of \opencypher queries to \rga and also give a more detailed listing of the compilation rules for the query language constructs in \autoref{table:mapping}. We follow a bottom-up approach to build the \rga expression.

\begin{enumerate}
\item Process each single query as follows and combine their result using the \uniontext operation. As the \uniontext operator is technically a binary operator, the \uniontext of more than two single queries are represented as a left-deep tree of \lstinline+UNION+ operators.

\item For each query part of a single query, denoted by $t$, the \rga tree built from the prefix of query parts up to---but not including---the current query part, process the current query part as follows.

\setlength\tabcolsep{3.6pt}
\begin{enumerate}[label=\arabic*.]
	\item A single pattern is turned left-to-right to a \getverticestext for the first vertex and a chain of \expandintext, \expandouttext or \expandbothtext operators for inbound, outbound or undirected relationships, respectively.
	\item Comma-separated patterns in a single \lstinline+MATCH+ are connected by \jointext.
	\item Append an \alldifferenttext operator for all edge variables that appear in the \lstinline+MATCH+ clause because of the non-repeating edges language rule.
	\item Process the \lstinline+WHERE+ subclause of a single \lstinline+MATCH+ clause.
	\item Several \lstinline+MATCH+ clauses are connected to a left-deep tree of \jointext.

	      For \lstinline+OPTIONAL MATCH+, \leftouterjointext is used instead of \jointext. In case there is a \lstinline+WHERE+ subclause, its condition becomes part of the join condition, \ie it will never filter on the input from previous \lstinline+MATCH+ clauses.
%If a single query is composed of more than one query parts, they are combined together using the \jointext operator.
	\item If there is a positive or negative pattern deferred from \lstinline+WHERE+ processing, append it as a \jointext or a combination of \leftouterjointext and \selectiontext operator filtering on no matches were found, respectively.
	\item If this is not the first query part, combine the curent query part with the \rga tree of the preceding query parts by appending a \jointext here. Its left operand will be $t$ and its right operand will be the \rga tree built so far from the current subquery.
	\item Append \groupingtext, if \lstinline+RETURN+ or \lstinline+WITH+ clause has grouping functions inside.
	\item Append a \projectiontext operator based on the \lstinline+RETURN+ or \lstinline+WITH+ clause. This operator will also handle the renaming (i.e. \lstinline+AS+).
	\item Append a \duplicateeliminationtext operator, if the \lstinline+RETURN+ or \lstinline+WITH+ clause has the \lstinline+DISTINCT+ modifier.
	\item Append a \selectiontext operator if \lstinline+WITH+ had the optional \lstinline+WHERE+ subclause.
\end{enumerate}

\end{enumerate}

\setlength\extrarowheight{2.5pt}
\setlength\tabcolsep{3.6pt}
\begin{table}[htbp]
\begin{adjustbox}{center}
\begin{tabular}{|l|l|l|c|}
	\hline
	\multicolumn{2}{|l|}{ \bf Language construct } & \bf Relational algebra expression & \\\hline\hline

	%\hline
	\multicolumn{4}{|l|}{Vertices and patterns. \lstinline|<p>p</p>| denotes a pattern that contains a vertex \lstinline|<<v>>|.} \\\cline{2-4}

	& \lstinline+(<<v>>)+ & $\getvertices{v}{}$ & \mapping{getvertices} \\\cline{2-4}

	& \lstinline+(<<v>>:<<l1>>:...:<<ln>>)+ & $\getvertices{v}{l1 \land \cdots \land ln}$ & \mapping{getverticeslabels} \\\cline{2-4}

	& \lstinline+<p>p</p>-[<<e>>:<<t1>>|...|<<tk>>]->(<<w>>)+ & $\expandout{v}{w}{}{e}{t1 \lor \cdots \lor tk}{1}{1} (p)$, where $e$ is an edge & \mapping{patterni} \\\cline{2-4}

	& \lstinline+<p>p</p><-[<<e>>:<<t1>>|...|<<tk>>]-(<<w>>)+ & $\expandin{v}{w}{}{e}{t1 \lor \cdots \lor tk}{1}{1} (p)$, where $e$ is an edge & \mapping{patternii} \\\cline{2-4}

	& \lstinline+<p>p</p><-[<<e>>:<<t1>>|...|<<tk>>]->(<<w>>)+ & $\expandboth{v}{w}{}{e}{t1 \lor \cdots \lor tk}{1}{1} (p)$, where $e$ is an edge & \mapping{patterniii} \\\cline{2-4}

	& \lstinline+<p>p</p>-[<<e>>*<<min>>..<<max>>]->(<<w>>)+ & $\expandout{v}{w}{}{e}{}{min}{max} (p)$, where $e$ is a list of edges& \mapping{patterniv} \\\cline{2-4}

	\hline \multicolumn{3}{|l|}{Combining and filtering pattern matches} & \\\cline{2-4}

	& \lstinline+MATCH <p>p1</p>, <p>p2</p>, ...+ &
	$\alldifferent{\atom{edges\ of\ p1,\ p2,\ \cdots}} \left( p1 \join p2 \join \cdots \right)$ & \mapping{matchi} \\\cline{2-4}

	& \breakable{
		\lstinline+MATCH <p>p1</p>+ \\
		\lstinline+MATCH <p>p2</p>+
	} &
	$\alldifferent{\atom{edges\ of\ p1}} \left(p1\right) \ \join\ \alldifferent{\atom{edges\ of\ p2}} \left(p2\right)$ & \mapping{matchii} \\\cline{2-4}

	& \breakable{
		\lstinline+OPTIONAL MATCH <p>p</p>+
	} & $ \{\tuple{}\} \ \leftouterjoin\ \alldifferent{\atom{edges\ of\ p}} \left(p\right)$ & \mapping{optionalmatchi} \\\cline{2-4}

	& \breakable{
		\lstinline+OPTIONAL MATCH <p>p</p> WHERE <p>condition</p>+
	} & $ \{\tuple{}\} \ \leftouterjoin_{\atom{condition}}\ \alldifferent{\atom{edges\ of\ p}} \left(p\right)$ & \mapping{optionalmatchwherei} \\\cline{2-4}

	& \breakable{
		\lstinline+[[r]] OPTIONAL MATCH <p>p</p>+
	} & $\alldifferent{\atom{edges\ of\ r}}\left(r\right) \ \leftouterjoin\ \alldifferent{\atom{edges\ of\ p}} \left(p\right)$ & \mapping{optionalmatchii} \\\cline{2-4}

	& \breakable{
		\lstinline+[[r]] WHERE <<condition>>+
	} & \breakable{$\selection{\atom{condition}}{\left( r \right)}$} & \mapping{wherei} \\\cline{2-4}

	& \breakable{
		\lstinline+[[r]] WHERE (<<v>>:<<l1>>:...:<<ln>>)+ \\
	} & \breakable{$\selection{\vertexlabelfunction(v) = \var{l1} \land \cdots \land \vertexlabelfunction(v) = \var{ln}} (r) $ } & \mapping{whereii} \\\cline{2-4}	

	& \breakable{
		\lstinline+[[r]] WHERE <p>p</p>+ \\
	} & \breakable{$ r \join p $ } & \mapping{whereiii} \\\cline{2-4}

	%& \breakable{
	%	\lstinline+[[r]] WHERE NOT <p>p</p>+ \\
	%} & \breakable{$ r \antijoin p $ } & \mapping{whereiv} \\\cline{2-4}

	\hline \multicolumn{4}{|l|}{Result and subresult operations. Rules for \lstinline+RETURN+ also apply to \lstinline+WITH+.} \\\cline{2-4}

	& \lstinline+[[r]] RETURN <<x1>> AS <<y1>>, ...+ & $\projection{\var{x1} \assign \var{y1}, \cdots}{} \left( r \right)$ & \mapping{returni} \\\cline{2-4}

	& \lstinline+[[r]] RETURN DISTINCT <<x1>> AS <<y1>>, ...+ & $\duplicateelimination\left(\projection{\var{x1} \assign \var{y1}, \cdots}{} \left( r \right)\right)$ & \mapping{returnii} \\\cline{2-4}

	& \lstinline+[[r]] RETURN <<x1>>, <<aggr>>(<<x2>>)+ & $\grouping{\var{x1}, \atom{aggr}(\var{x2})}{\var{x1}}{\left( r \right)}$ (see \autoref{sec:aggregation}) & \mapping{returniii} \\\cline{2-4}

	& \breakable{
		\lstinline+[[r]] WITH <<x1>>+ \\
		\lstinline+[[s]] RETURN <<x2>>+
	} & \breakable{$\projection{\var{x2}}{}\Big(\big(\projection{\var{x1}}{} \left( r \right) \big) \join s\Big)$ } & \mapping{returniv} \\\cline{2-4}

	\hline \multicolumn{4}{|l|}{Unwinding and list operations} \\\cline{2-4}

	& \lstinline+[[r]] UNWIND <<xs>> AS <<x>>+ & $\unwind{\var{xs} \assign \var{x}}{\left( r \right)}$ & \mapping{unwind} \\\cline{2-4}

	& \lstinline+[[r]] ORDER BY <<x1>> ASC, <<x2>> DESC, ...+ & $\sort{\asc \var{x1}, \desc \var{x2}, \cdots}{\left( r \right)}$ & \mapping{orderby} \\\cline{2-4}

	& \lstinline+[[r]] SKIP <<s>> LIMIT <<l>>+ & $\topp{\var{l}}{\var{s}}(r)$
	& \mapping{top} \\\cline{2-4}

	\hline \multicolumn{4}{|l|}{Combining results } \\\cline{2-4}

	& \lstinline+[[r]] UNION [[s]] + & $r \union s$ & \mapping{union} \\\cline{2-4}

	& \lstinline+[[r]] UNION ALL [[s]] + & $r \bagunion s$ & \mapping{unionall} \\\hline
\end{tabular}
\end{adjustbox}
\caption{Mapping from \opencypher constructs to relational algebra. Variables, labels, types and literals are typeset as \lstinline|<<v>>|. The notation \lstinline|<p>p</p>| represents patterns resulting in a relation $p$, while \lstinline|[[r]]| denotes previous query fragment resulting in a relation $r$. To avoid confusion with the ``\lstinline+..+'' language construct (used for ranges), we use \lstinline+...+ to denote omitted query fragments.}
\label{table:mapping}
\end{table}

\subsection{Summary and Limitations}

In this section, we presented a mapping that allows us to express the example queries of \autoref{sec:language-constructs} in graph relational algebra. We extended relational algebra by adapting operators ($\getverticesop$, $\expandoutop$, $\sortop$, $\topop$), precisely specifying grouping semantics ($\groupingop$) and defining the all-different operator ($\alldifferentop$). Finally, we proposed an algorithm for compiling \opencypher graph queries to graph relational algebra.

Our mapping does not completely cover the \opencypher language. As discussed in \autoref{sec:opencypher}, some constructs are defined as legacy and thus were omitted. The current formalisation does not include expressions (\eg conditions in selections) and maps. Compiling data manipulation operations (such as \mbox{\lstinline+CREATE+}, \mbox{\lstinline+DELETE+}, \mbox{\lstinline+SET+,} and \mbox{\lstinline+MERGE+}) to relational algebra is also subject of future work.

% !TeX spellcheck = en_GB
% !TeX encoding = UTF-8
\section{Related Work}
\label{sec:related-work}

\subsubsection{Property graph data models.} The TinkerPop framework aims to provide a standard data model for property graphs, along with Gremlin, a high-level imperative graph traversal language~\cite{Rodriguez:2015:GGT:2815072.2815073} and the Gremlin Structure API, a low-level programming interface.

\vspace{-2ex}

\subsubsection{EMF.} The Eclipse Modeling Framework is an object-oriented modelling framework widely used in model-driven engineering.
Henshin~\cite{DBLP:conf/models/ArendtBJKT10} provides a visual language for defining patterns, while Epsilon~\cite{DBLP:conf/icmt/KolovosPP08} and \viatraquery~\cite{DBLP:conf/models/BergmannHRVBBO10} provide high-level declarative (textual) query languages, the Epsilon Pattern Language and the \vql.

\vspace{-2ex}

\subsubsection{SPARQL.} Widely used in semantic technologies, \sparql is a standardised declarative graph pattern language for querying RDF~\cite{RDF} graphs. SPARQL bears close similarity to Cypher queries, but targets a different data model and requires users to specify the query as triples instead of graph vertices/edges%. A formal definition of the language is given in
~\cite{DBLP:journals/tods/PerezAG09}. \mbox{G-SPARQL}~\cite{DBLP:conf/cikm/SakrEH12} extended the SPARQL language for attributed graphs, resulting in a language with an expressive power similar to \opencypher.

\vspace{-2ex}

\lstset{language=}
\subsubsection{SQL.} In general, relational databases offer limited support for graph queries: recursive queries are supported by \mbox{PostgreSQL} using the \lstinline+WITH RECURSIVE+ keyword and by the Oracle Database using the \lstinline+CONNECT BY+ keyword. Graph queries are supported in the \saphana %Graph Scale-Out Extension 
prototype~\cite{DBLP:conf/btw/RudolfPBL13}, through a SQL-based language~\cite{DBLP:conf/gg/KrauseJDSKN16}.

\vspace{-2ex}

\subsubsection{Cypher.} Due to its novelty, there are only a few research works on the formalisation of (open)Cypher. The authors of~\cite{DBLP:conf/edbt/HolschG16} defined \emph{graph relations} and introduced the \mbox{\textsc{GetNodes}}, \mbox{\textsc{ExpandIn}} and \mbox{\textsc{ExpandOut}} operators. While their work focused on optimisation transformations, this paper aims to provides a more complete and systematic mapping from openCypher to relational algebra.

In~\cite{DBLP:conf/sigmod/Junghanns17}, graph queries were defined in a Cypher-like language and evaluated on Apache Flink. % and evaluated in the Apache Flink-based \textsc{Gradoop} framework.
However, formalisation of the queries was not discussed in detail.

\vspace{-2ex}

\subsubsection{Comparison of graph query frameworks.} Previously, we published the Train Benchmark, a framework for comparing graph query frameworks across different technological spaces, such as property graphs, EMF, RDF and SQL~\cite{TrainBenchmarkSOSYM}.

\vspace{-2ex}

% !TeX spellcheck = en_GB
% !TeX encoding = UTF-8
\section{Conclusion and Future Work}
\label{sec:conclusion}

In this paper, we presented a formal specification for a subset of the \opencypher query language. This provides the theoretical foundations to use \opencypher as a language for graph query engines.

As future work, we plan to provide a formalisation based on graph-specific theoretical query frameworks, such as~\cite{DBLP:journals/jacm/LibkinMV16}. We will also give the formal specification of the operators for incremental query evaluation, which requires the definition of \emph{maintenance operations} that keep the result in sync with the latest set of changes~\cite{PerPolQueryOptimization}. Our long-term research objective is to design an \opencypher-compatible \emph{distributed, incremental graph query engine}~\cite{DBLP:conf/models/SzarnyasIRHBV14}.\footnote{Our prototype, \emph{ingraph}, is available at: \url{http://docs.inf.mit.bme.hu/ingraph/}}

% !TeX spellcheck = en_GB
% !TeX encoding = UTF-8
\section*{Acknowledgements}

\vspace{-1ex}

G\'abor Sz\'arnyas and D\'aniel Varr\'o were supported by the MTA-BME Lend\"ulet Research Group on Cyber-Physical Systems and the NSERC RGPIN-04573-16 project. The authors would like to thank G\'abor Bergmann and J\'anos Maginecz for their comments on the draft of this paper.

\vspace{-2ex}

\bibliographystyle{abbrv}
\bibliography{ms}

\begin{thebibliography}{10}

\bibitem{DBLP:conf/models/ArendtBJKT10}
T.~Arendt et~al.
\newblock Henshin: Advanced concepts and tools for in-place {EMF} model
  transformations.
\newblock In {\em {MODELS}}, pages 121--135, 2010.

\bibitem{DBLP:conf/models/BergmannHRVBBO10}
G.~Bergmann et~al.
\newblock Incremental evaluation of model queries over {EMF} models.
\newblock In {\em {MODELS}}, pages 76--90, 2010.

\bibitem{DBLP:conf/dlog/BotoevaCCRX16}
E.~Botoeva et~al.
\newblock {OBDA} beyond relational {DBs}: {A} study for {MongoDB}.
\newblock In {\em Proceedings of the 29th International Workshop on Description
  Logics}, 2016.

\bibitem{DBLP:books/daglib/0006733}
R.~Elmasri and S.~B. Navathe.
\newblock {\em Fundamentals of Database Systems}.
\newblock Addison-Wesley-Longman, 3rd edition, 2000.

\bibitem{DBLP:conf/sigmod/ErlingALCGPPB15}
O.~Erling et~al.
\newblock The {LDBC} {S}ocial {N}etwork {B}enchmark: Interactive workload.
\newblock In {\em {SIGMOD}}, pages 619--630, 2015.

\bibitem{DBLP:books/daglib/0020812}
H.~Garcia{-}Molina, J.~D. Ullman, and J.~Widom.
\newblock {\em Database systems -- {T}he complete book}.
\newblock Pearson Education, 2nd edition, 2009.

\bibitem{DBLP:conf/edbt/HolschG16}
J.~H{\"{o}}lsch and M.~Grossniklaus.
\newblock An algebra and equivalences to transform graph patterns in {N}eo4j.
\newblock In {\em {GraphQ} at {EDBT/ICDT}}, 2016.

\bibitem{DBLP:conf/sigmod/Junghanns17}
M.~Junghanns et~al.
\newblock Cypher-based graph pattern matching in {Gradoop}.
\newblock In {\em {GRADES} at {SIGMOD}}, 2017.

\bibitem{DBLP:conf/icmt/KolovosPP08}
D.~S. Kolovos et~al.
\newblock The {E}psilon transformation language.
\newblock In {\em {ICMT}}, 2008.

\bibitem{DBLP:conf/gg/KrauseJDSKN16}
C.~Krause et~al.
\newblock An {SQL}-based query language and engine for graph pattern matching.
\newblock In {\em ICGT}, pages 153--169, 2016.

\bibitem{DBLP:conf/sigmod/LiCIS05}
C.~Li, K.~C. Chang, I.~F. Ilyas, and S.~Song.
\newblock Rank{SQL}: Query algebra and optimization for relational top-k
  queries.
\newblock In {\em {SIGMOD}}, pages 131--142, 2005.

\bibitem{DBLP:journals/jacm/LibkinMV16}
L.~Libkin et~al.
\newblock Querying graphs with data.
\newblock {\em J. {ACM}}, 63(2):14:1--14:53, 2016.

\bibitem{openCypher}
{Neo Technology}.
\newblock open{C}ypher project.
\newblock \url{http://www.opencypher.org/}, 2017.

\bibitem{DBLP:journals/tods/PerezAG09}
J.~P{\'{e}}rez et~al.
\newblock Semantics and complexity of {SPARQL}.
\newblock {\em {ACM} {TODS}}, 34(3), 2009.

\bibitem{CollectivelyGeneratedModel}
M.~A. Rodriguez.
\newblock A collectively generated model of the world.
\newblock In {\em Collective intelligence: creating a prosperous world at
  peace}, pages 261--264. 2008.

\bibitem{Rodriguez:2015:GGT:2815072.2815073}
M.~A. Rodriguez.
\newblock The {G}remlin graph traversal machine and language (invited talk).
\newblock In {\em DBPL}, pages 1--10, 2015.

\bibitem{DBLP:books/igi/Sakr11/RodriguezN11}
M.~A. Rodriguez and P.~Neubauer.
\newblock The graph traversal pattern.
\newblock In {\em Graph Data Management: Techniques and Applications}, pages
  29--46. 2011.

\bibitem{DBLP:conf/btw/RudolfPBL13}
M.~Rudolf et~al.
\newblock The graph story of the {SAP} {HANA} database.
\newblock In {\em BTW}, 2013.

\bibitem{DBLP:conf/cikm/SakrEH12}
S.~Sakr, S.~Elnikety, and Y.~He.
\newblock {G-SPARQL:} a hybrid engine for querying large attributed graphs.
\newblock In {\em {CIKM}}, pages 335--344, 2012.

\bibitem{DBLP:conf/models/SzarnyasIRHBV14}
G.~Sz{\'{a}}rnyas et~al.
\newblock Inc{Q}uery-{D}: {A} distributed incremental model query framework in
  the cloud.
\newblock In {\em {MODELS}}, pages 653--669, 2014.

\bibitem{TrainBenchmarkSOSYM}
G.~Sz{\'a}rnyas et~al.
\newblock The {T}rain {B}enchmark: Cross-technology performance evaluation of
  continuous model validation.
\newblock {\em Softw. Syst. Model.}, 2017.

\bibitem{PerPolQueryOptimization}
G.~Sz{\'a}rnyas, J.~Maginecz, and D.~Varr{\'o}.
\newblock Evaluation of optimization strategies for incremental graph queries.
\newblock {\em Periodica Polytechnica, EECS}, 2017.

\bibitem{openCypherReport}
G.~Sz\'arnyas and J.~Marton.
\newblock Formalisation of {openCypher} queries in relational algebra.
\newblock Technical report, Budapest University of Technology and Economics,
  2017.
\newblock \url{http://hdl.handle.net/10890/5395}.

\bibitem{RDF}
{W3C}.
\newblock {R}esource {D}escription {F}ramework.
\newblock \url{https://www.w3.org/RDF/}, 2014.

\end{thebibliography}

\end{document}